\documentclass[pre,showpacs,showkeys,preprintnumbers,amsmath,amssymb,superscriptaddress,onecolumn]{revtex4-2}
\usepackage{graphicx}
\usepackage{amsfonts}
\usepackage{url}
\usepackage{epstopdf}
\usepackage{color}
\usepackage{tikz-cd}
\usepackage{comment}
\usepackage{bm}
\usepackage[colorlinks=true,linkcolor=blue,citecolor=red]{hyperref}%

\def\X{\bm{X}}
\def\x{\bm{x}}
\def\erfc{\mathrm{erfc}}
\def\P{\mathbb{P}}
\def\R{{\mathbb R}}
\def\pa{\partial\Omega}

\def\N{\mathcal N}

\def\T{{\mathcal T}}
\def\E{{\mathbb E}}
\def\TT{\mathbb{T}}
\def\C{\mathbb{C}}

\def\erfcx{\mathrm{erfcx}}
\def\arccosh{\mathrm{arccosh}}

\begin{document}

\title{Adsorption and permeation events in molecular diffusion}

\author{Denis~S.~Grebenkov}
 \email{denis.grebenkov@polytechnique.edu}
\affiliation{
Laboratoire de Physique de la Mati\`{e}re Condens\'{e}e, \\
CNRS -- Ecole Polytechnique, Institut Polytechnique de Paris, 91120 Palaiseau, France}
\affiliation{
Centre de recherches math\'ematiques (CRM), Universit\'e de Montr\'eal, \\
6128 succ Centre-Ville, Montr\'eal QC H3C 3J7, Canada}

\date{\today}

\begin{abstract}
How many times a diffusing molecule can permeate across a membrane or
be adsorbed on a substrate?  We employ the encounter-based approach to
find the statistics of adsorption or permeation events for molecular
diffusion in a general confining medium.  Various features of these
statistics are illustrated for two practically relevant cases of a
flat boundary and a spherical confinement.  Some applications of these
fundamental results are discussed.
\end{abstract}




\keywords{diffusion; surface reaction; permeation; heterogeneous catalysis;
confinement; geometric complexity; biochemistry; reversible reactions;
encounter-based approach; Brownian motion}

\maketitle

\section{Introduction}
\label{sec:intro}

Many chemical and biological phenomena are governed by molecular
diffusion \cite{Redner,House,Schuss,Metzler,Masoliver,Oshanin,Dagdug},
with examples ranging from oxygen capture by blood in lungs and
placentas \cite{Weibel,Grebenkov05,Serov16,Sapoval21} to proteins
searching for enzymes or specific sites on a DNA chain
\cite{Benichou11,Kolomeisky11,Sheinman12,Bressloff13}.  In such
diffusion-controlled reactions, the diffusion-driven mixture of
reacting molecules or their diffusive transport to immobile reactive
patches are relevant and often limiting factors in the overall
chemical kinetics
\cite{North66,Wilemski73,Calef83,Berg85,Rice85,Grebenkov23f}.  In
addition, the diffusing molecules can reversibly bind to other
molecules or adsorb/desorb on a substrate
\cite{Goodrich54,Mysels82,Agmon84,Agmon89,Agmon90,Kim99,Prustel13,Grebenkov19g,Scher23f}.
For instance, reversible binding of calcium ions to buffer molecules
controls the cascade of biochemical reactions in the presynaptic
bouton and thus signal transduction between neurons
\cite{Sala90,Neher08,Holcman13,Guerrier18,Reva21}.  Moreover, if the
confining medium is compartmentalized by semi-permeable membranes, the
molecules can permeate across them and thus explore different
compartments
\cite{Dick64,Crick70,Michel99,Sapoval02,Novikov11,Kay22,Bressloff23a,Bressloff23b}.
While the theory of such reversible reactions is rather well developed
(see reviews \cite{Scher23f,Grebenkov23f} and references therein), the
statistics of adsorption and permeation events remains poorly
understood.  How many times a single molecule would cross the membrane
or would experience adsorption/desorption up to a given time or before
its escape, degradation, passivation or ultimate transformation?  This
fundamental question can shed light onto the intricate diffusive
dynamics of molecules in complex media and help to design more
efficient mechanisms and protocols for targeted drug delivery
\cite{Siepmann08,Carr18}.

Different mathematical tools were employed to study the dynamics of
reversible diffusion-controlled reactions.  Some theoretical models
substitute the spatial dynamics of a molecule by a random walk on a
lattice
\cite{Rubin82,Haus87,Nicolis01,Grebenkov03,Moreau04,Schmit09,Abad15,Giuggioli20},
in which case the number of adsorption or permeation events can be
easily introduced.  However, the obtained statistics depend on the
lattice size, and finding its asymptotic behavior presents a difficult
task.  Moreover, the related analytical computations rely on
combinatorial analysis and are generally accessible only for few
simple geometric settings.  As a consequence, one usually employs
another framework, in which the dynamics of a molecule is described as
Brownian motion, restricted by reflecting walls of the confinement.
In this framework, one can consider either the macroscopic
concentration of molecules, or the diffusive propagator characterizing
the random position of a single molecule.  In both cases, theoretical
or numerical analysis focuses on the diffusion equation with
appropriate boundary conditions, for which various mathematical tools
are available.  In particular, when the target is small or the
confinement is large, matched asymptotic analysis and scaling
arguments yield an accurate description of first-passage times (see
\cite{Singer06,Benichou08,Pillay10,Benichou14,Holcman14,Grebenkov16,Bressloff22b,Grebenkov22c}
and references therein).  At the same time, the very definition of an
adsorption or permeation event is getting more sophisticated.  In
fact, when Brownian motion arrives onto a reflecting boundary, it
exhibits infinitely many reflections within an infinitely short time
period \cite{Morters}.  This paradoxical property is a mathematical
consequence of self-similarity of Brownian motion at all scales.
Accordingly, if an adsorption or permeation event is initiated upon
the first arrival onto the boundary, the molecule cannot leave it to
resume its motion in the bulk, as it will be immediately re-adsorbed.
To overcome this technical difficulty, one can realize desorption or
permeation as a jump from the boundary into the bulk to a small
distance $a > 0$ \cite{Grebenkov07,Benichou10}, that may represent a
finite thickness of a reactive surface layer, the thickness of a
permeable membrane, or the short range of interactions between the
molecule and the boundary.  Alternatively, one can introduce a finite
reactivity, permeability, or binding affinity $\kappa$ of the molecule
to the boundary so the molecule has a finite probability to adsorb or
permeate at each arrival onto the boundary
\cite{Collins49,Sano79,Brownstein79,Sano81,Agmon88,Sapoval94,Erban07,Galanti16,Grebenkov19c,Grebenkov20f,Piazza22}.
Both options yield similar results when $a$ is small, but the second
option has the advantage of preserving continuity of random
trajectories.

Recently, a more rigorous formulation of adsorption and permeation
events was elaborated
\cite{Grebenkov20,Grebenkov20c,Bressloff22c,Grebenkov23a}.  This
so-called encounter-based approach relies on the boundary local time
$\ell_t$, a mathematical process that represents the rescaled number
of encounters between the molecule and the boundary up to time $t$.
This formulation provides a rigorous foundation for the earlier works
based on the finite reactivity $\kappa$ and allows one to study much
more general binding and permeation mechanisms.  This approach turns
out to be particularly suitable for describing a sequence of diffusive
explorations near the boundary after each failed adsorption/permeation
attempt.  In this paper, we employ the encounter-based approach to
obtain the exact statistics of adsorption or permeation events.  As
permeation across the membrane involves diffusion in distinct
compartments, the related analysis is much more difficult.  To avoid
the related technicalities and to keep the presentation to an
accessible level, we focus mostly on the adsorption events.  After
deriving the general solution for the statistics of adsorption events
(Sec. \ref{sec:adsorption}) and permeation events
(Sec. \ref{sec:permeation}), we present two relevant examples of
adsorptions on a flat boundary and on a spherical boundary
(Sec. \ref{sec:examples}).  In both cases, we manage to get rather
explicit formulas to illustrate how the statistics of adsorption
events depend on other parameters.  Section
\ref{sec:conclusion} concludes the paper with the summary of main
findings and future perspectives.

\section{Statistics of adsorption events}
\label{sec:adsorption}

We consider molecular diffusion in a Euclidean domain $\Omega\subset
\R^d$ whose boundary $\pa = \Gamma \cup \pa_0$ is split into two
disjoint parts: an adsorbing surface $\Gamma$ and a reflecting inert
boundary $\pa_0$.  For a molecule started from a point $\x_0 \in
\Omega$, we aim at characterizing the statistics of the number $\N_t$
of adsorptions on $\Gamma$ up to time $t$.  This problem resembles the
escape problem for a sticky particle \cite{Scher23,Scher24}.  The
difference here is that we focus on a fixed time $t$, whereas the
escape problem corresponded to $\N_\TT$, with $\TT$ being the
first-escape time.  For a better understanding of this statistics, we
assume that the adsorption events are infinitely short, i.e., an
adsorbed molecule is immediately released and resumes its diffusion
until the next adsorption.  Note that random durations of adsorption
events can be easily incorporated into the proposed formalism (see the
paragraph after Eq. (\ref{eq:Np2})).

\begin{figure}
\begin{center}
\includegraphics[width=0.44\textwidth]{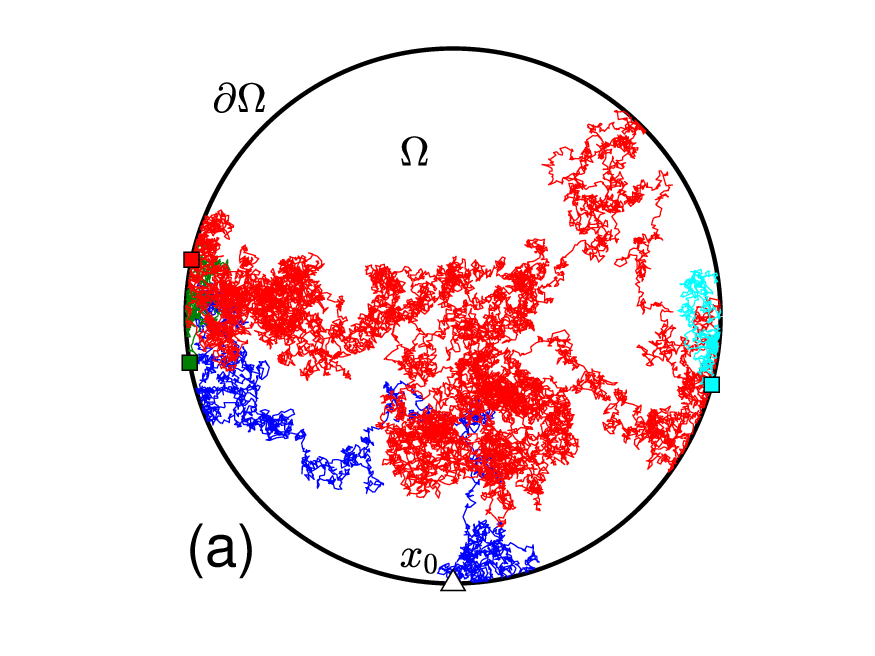} 
\includegraphics[width=0.55\textwidth]{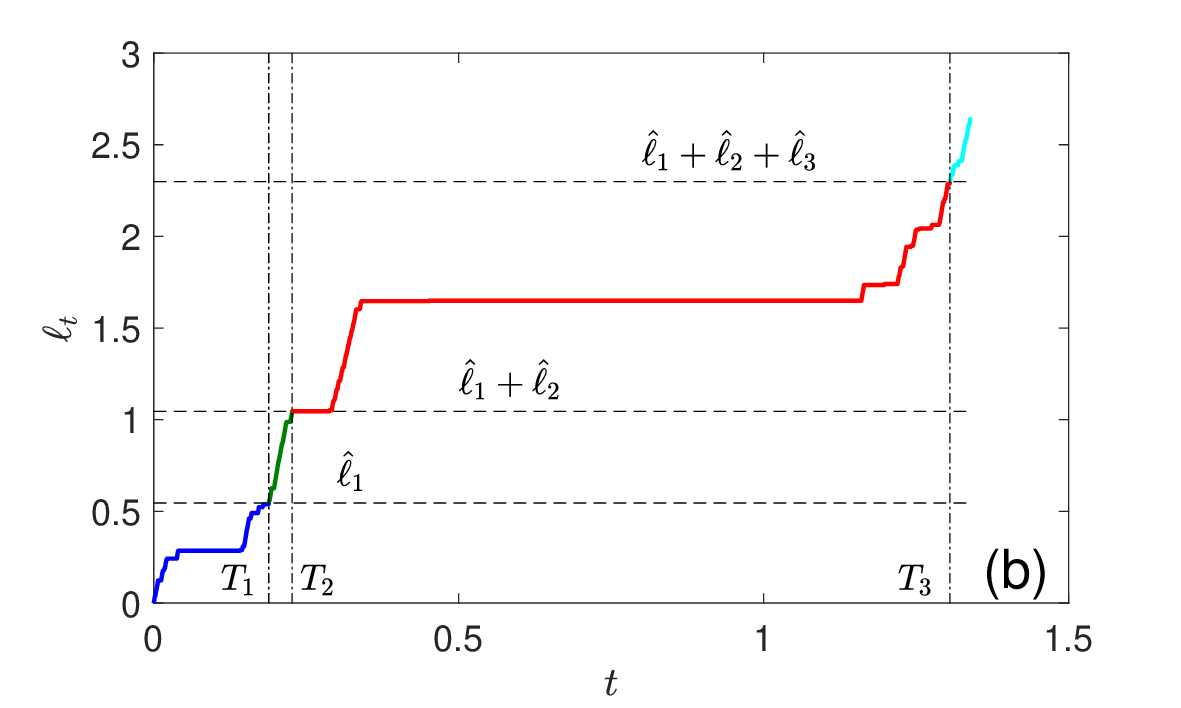} 
\end{center}
\caption{
{\bf (a)} A simulated trajectory of a molecule diffusing inside a disk
$\Omega$ with an adsorbing circular boundary $\Gamma = \pa$.  A
triangle indicates the starting point $\x_0$, while three filled
squares show three positions, at which the molecule was adsorbed to
the boundary.  {\bf (b)} The associated boundary local time $\ell_t$
(obtained along the same simulation) crosses three random thresholds
$\hat{\ell}_1$, $\hat{\ell}_1 + \hat{\ell}_2$, and $\hat{\ell}_1 +
\hat{\ell}_2 + \hat{\ell}_3$ (horizontal lines) at random times $\T_1$,
$\T_2$, and $\T_3$ (vertical lines).  Each such crossing corresponds
to an adsorption on the boundary.  Colors distinguish successive time
periods between adsorption events.}
\label{fig:scheme}
\end{figure}

First of all, one needs to formulate the adsorption/desorption
mechanism.  We follow the encounter-based approach, in which the
adsorption event occurs when the number of encounters between the
molecule and the adsorbing boundary $\Gamma$ exceeds some threshold
\cite{Grebenkov07,Grebenkov19b}.  In more rigorous terms, the random
adsorption time $\T$ is defined as $\T = \inf\{ t > 0~:~ \ell_t >
\hat{\ell}\}$, where $\ell_t$ is the boundary local time on $\Gamma$
(i.e., a rescaled number of encounters with $\Gamma$ up to time $t$)
and $\hat{\ell}$ is the random threshold with a given distribution
\cite{Grebenkov20}.  For instance, if $\hat{\ell}$ obeys an
exponential law, $\P\{ \hat{\ell} > \ell\} = e^{-q\ell}$ with some $q
> 0$, this mechanism describes adsorption to the boundary with a
constant reactivity $\kappa = qD$, governed by Robin boundary
condition \cite{Grebenkov20,Grebenkov23a}.  In the following, we will
treat more general adsorption mechanisms characterized by a given
probability law $\Psi(\ell) =\P\{ \hat{\ell} > \ell\}$ for the
threshold $\hat{\ell}$, and then discuss the special case of a
constant reactivity.
In this framework, the desorbed molecule can be released from the
adsorption point on the boundary $\Gamma$, thus preserving the
continuity of the trajectory.  As each desorption event is associated
to resetting of the boundary local time $\ell_t$ to zero, the molecule
needs to encounter the boundary a number of times before being
re-adsorbed, resulting in a well-defined adsorption/desorption
kinetics.  Instead of resetting the boundary local time $\ell_t$, one
can equivalently introduce a sequence of independent thresholds
$\hat{\ell}_1, \hat{\ell}_2, \ldots, \hat{\ell}_n$ to define a
sequence of adsorption times,
\begin{equation}
\T_n = \inf\{ t > 0 ~:~ \ell_t > \hat{\ell}_1 + \ldots + \hat{\ell}_n\}   \qquad (n = 1,2,\ldots).
\end{equation}
Figuratively speaking, thresholds $\hat{\ell}_1$,
$\hat{\ell}_1+\hat{\ell}_2$, etc. can be interpreted as ``milestones''
on a road, whereas $\T_1$, $\T_2$, etc. are the successive crossing
times of these milestones (Fig. \ref{fig:scheme}).  We are interested
in describing the statistics of the number of adsorptions up to time
$t$:
\begin{equation}
\N_t = \max\{n\geq 0 ~:~ \T_n < t\}
\end{equation}
(with $\T_0 = 0$), i.e., the number of adsorption times $\T_n$ that do
not exceed $t$.  Since $\N_t$ is a discrete stochastic process, we can
fully characterize it by finding the probabilities $Q_n(t|\x_0) = \P\{
\N_t = n\}$, with $n = 0,1,2,\ldots$.

While our main emphasis is on the statistics of $\N_t$ for a fixed
time $t$, another important statistics can be considered.  For this
extension, we assume that the diffusing molecule has a finite random
lifetime $\tau$ \cite{Yuste13,Meerson15,Grebenkov17f}.  This lifetime
can account for eventual degradation or passivation of the molecule,
or its irreversible transformation into another (inactive) molecule.
What is the statistics of the random variable $\N_\tau$, i.e., the
number of adsorption events realized during the lifetime?  If the
random variable $\tau$ is independent of the dynamics, the statistics
of $\N_\tau$ can be obtained by averaging the statistics of $\N_t$
with the PDF $\rho(t)$ of $\tau$:
\begin{equation}
Q_n(\x_0) = \E\{ Q_n(\tau|\x_0)\} = \int\limits_0^\infty dt \, \rho(t) \, Q_n(t|\x_0) .
\end{equation}
In the most common case of an exponentially distributed lifetime,
$\rho(t) = p e^{-pt}$ (with the rate $p > 0$ or mean lifetime $1/p$),
the right-hand side of this expression is proportional to the Laplace
transform of $Q_n(t|\x_0)$,
\begin{equation}  \label{eq:Qn_lifetime}
Q_n(\x_0) = p \int\limits_0^\infty dt \, e^{-pt} \, Q_n(t|\x_0) = p\, \tilde{Q}_n(p|\x_0) ,
\end{equation}
where tilde denotes the Laplace transform with respect to time.  In
the next subsection, we will derive a rather explicit form for the
Laplace transform $\tilde{Q}_n(p|\x_0)$ so that the statistics of such
$\N_\tau$ is actually much simpler than that of $\N_t$.

\subsection{Encounter-based approach}

In order to compute $Q_n(t|\x_0)$, we employ the encounter-based
approach \cite{Grebenkov20}.  Following this reference, we introduce
the generalized propagator $G(\x,t|\x_0)$, i.e., the probability
density for a molecule started from $\x_0\in\Omega$ at time $0$ to be
found in a vicinity of $\x\in\Omega$ at time $t$ without being
adsorbed up to $t$.  Since the no adsorption condition corresponds to
$\ell_t > \hat{\ell}$, the generalized propagator can be written as
\begin{equation}
G(\x,t|\x_0) = \int\limits_0^\infty d\ell \, \Psi(\ell) \, P(\x,\ell,t|\x_0),
\end{equation}
where $P(\x,\ell,t|\x_0)$ is the encounter propagator, i.e., the joint
probability density of finding the molecule at point $\x$ with the
boundary local time $\ell$ at time $t$, given that it started from
$\x_0$.  The generalized propagator determines the probability flux
density, $j(\x,t|\x_0) = - D\partial_n G(\x,t|\x_0)|_{\Gamma}$, where
$\partial_n$ is the normal derivative oriented outwards the domain
$\Omega$.  This is the joint probability density of the adsorption
time $\T$ and the position $\X_\T$ of the adsorption event on the
boundary $\Gamma$.  The Markovian property of the diffusive dynamics
allows us to write:
\begin{align}
Q_0(t|\x_0) & = S(t|\x_0) , \\  \nonumber 
Q_n(t|\x_0) & = \int\limits_{\Gamma} d\x_1 \int\limits_0^t dt_1 \, j(\x_1,t_1|\x_0) 
 \int\limits_{\Gamma} d\x_2 \int\limits_{t_1}^t dt_2 \, j(\x_2,t_2-t_1|\x_1) \ldots \\   \label{eq:Pnt}
& \times \int\limits_{\Gamma} d\x_n \int\limits_{t_{n-1}}^t dt_n \, j(\x_n,t_n-t_{n-1}|\x_{n-1}) S(t-t_n|\x_n),
\end{align}
where 
\begin{equation}  \label{eq:Spsi_def}
S(t|\x_0) = \int\limits_t^\infty dt' \int\limits_{\Gamma} d\x \, j(\x,t'|\x_0)
\end{equation}
is the survival probability of the molecule up to time $t$, i.e., the
probability of no adsorption up to time $t$.  Indeed, the expression
(\ref{eq:Pnt}) describes the molecule that was adsorbed for the first
time at time $t_1$ at point $\x_1 \in \Gamma$, then desorbed
immediately, diffused in $\Omega$, re-adsorbed at time $t_2$ at point
$\x_2\in \Gamma$, etc., until the last adsorption at time $t_n$ at
point $\x_n\in \Gamma$.  The Laplace transform allows one to get rid
off convolutions in time, yielding
\begin{align}  \nonumber
\tilde{Q}_n(p|\x_0) & = \int\limits_0^\infty dt\, e^{-pt} \, Q_n(t|\x_0) \\   \label{eq:P_auxil1}
& = \int\limits_{\Gamma} d\x_1 \, \tilde{j}(\x_1,p|\x_0) \int\limits_{\Gamma} d\x_2 \, \tilde{j}(\x_2,p|\x_1) \ldots   
 \int\limits_{\Gamma} d\x_n \, \tilde{j}(\x_n,p|\x_{n-1})\, \tilde{S}(p|\x_n).
\end{align}
However, convolutions in space (i.e., the integrals over $\x_1,
\ldots,\x_n \in \Gamma$) are difficult to deal with by conventional
tools such as spectral expansions over Laplacian eigenfunctions
\cite{Redner}.  In contrast, one can rely on spectral expansions based
on the Steklov eigenfunctions \cite{Levitin}.  Indeed, the Laplace
transform of the encounter propagator reads \cite{Grebenkov20}
\begin{align}  \label{eq:P_spectral}
\tilde{P}(\x,\ell,p|\x_0) & = \tilde{G}_\infty(\x,p|\x_0) \delta(\ell)  
 + \frac{1}{D} \sum\limits_{k=0}^\infty [V_k^{(p)}(\x_0)]^* V_k^{(p)}(\x) e^{-\mu_k^{(p)}\ell} ,
\end{align}
where $\delta(\ell)$ is the Dirac distribution,
$\tilde{G}_\infty(\x,p|\x_0)$ is the Green's function of the modified
Helmholtz equation with mixed Dirichlet-Neumann boundary conditions,
\begin{subequations}
\begin{align}
(p - D\Delta) \tilde{G}_\infty(\x,p|\x_0) &= \delta(\x-\x_0) \quad (\x\in\Omega), \\
 \tilde{G}_\infty(\x,p|\x_0) &= 0 \quad (\x\in \Gamma), \\
\partial_n \tilde{G}_\infty(\x,p|\x_0) &= 0 \quad (\x\in \pa_0),
\end{align}
\end{subequations}
and $\mu_k^{(p)}$ and $V_k^{(p)}$ are the eigenvalues and
eigenfunctions of the Steklov-Neumann spectral problem:
\begin{subequations}  \label{eq:Steklov}
\begin{align}
(p - D\Delta) V_k^{(p)} &= 0 \qquad \textrm{in}~\Omega ,  \\
\partial_n V_k^{(p)} &= \mu_k^{(p)} V_k^{(p)} \qquad \textrm{on} ~ \Gamma , \\
\partial_n V_k^{(p)} &= 0 \qquad \textrm{on} ~ \pa_0 . 
\end{align}
\end{subequations}
When the boundary $\Gamma$ is bounded, the spectrum of the Steklov
problem is discrete, the eigenvalues can be ordered in an increasing
sequence, $0 \leq \mu_0^{(p)} \leq \mu_1^{(p)} \leq \ldots \nearrow
+\infty$, whereas the restrictions $v_k^{(p)} = V_k^{(p)}|_{\Gamma}$
of Steklov eigenfunctions onto the adsorbing boundary $\Gamma$ form a
complete orthonormal basis of the space $L^2(\Gamma)$ of
square-integrable functions \cite{Levitin,Girouard17,Colbois24}.

After simplifications, the spectral expansion (\ref{eq:P_spectral})
implies
\begin{equation}  \label{eq:j_spectral}
\tilde{j}(\x,p|\x_0) = \sum\limits_{k=0}^\infty [V_k^{(p)}(\x_0)]^* v_k^{(p)}(\x)  \, \Upsilon(\mu_k^{(p)}),
\end{equation}
where 
\begin{equation}
\Upsilon(\mu) = \E\{ e^{-\mu \hat{\ell}} \} = \int\limits_0^\infty d\ell \, e^{-\mu \ell} \, (-\partial_\ell \Psi(\ell))
\end{equation}
is the moment-generating function of the threshold $\hat{\ell}$, with
$(-\partial_\ell \Psi(\ell))$ being its probability density function
(PDF).  In particular, Eq. (\ref{eq:Spsi_def}) implies
\begin{equation}  \label{eq:S_spectral}
\tilde{S}(p|\x_0) = \frac{1}{p} - \frac{1}{p} \sum\limits_{k=0}^\infty   \Upsilon(\mu_k^{(p)}) c_k^{(p)}(\x_0),
\end{equation}
where
\begin{equation}  \label{eq:ck_def}
c_k^{(p)}(\x_0) = [V_k^{(p)}(\x_0)]^*  \int\limits_{\Gamma} v_k^{(p)}(\x)  .
\end{equation}
Substituting the spectral expansions (\ref{eq:j_spectral},
\ref{eq:S_spectral}) into Eq. (\ref{eq:P_auxil1}), one can use the
orthogonality of $\{v_k^{(p)}\}$ to get
\begin{subequations}  \label{eq:Qn_tilde}
\begin{align}
\tilde{Q}_0(p|\x_0) & = \frac{1}{p} - \tilde{R}_1(p|\x_0)   , \\
\tilde{Q}_n(p|\x_0) & = \tilde{R}_n(p|\x_0) - \tilde{R}_{n+1}(p|\x_0) , 
\end{align}
\end{subequations}
where
\begin{equation}  \label{eq:Rtilde_def}
\tilde{R}_n(p|\x_0) = \frac{1}{p} \sum\limits_{k=0}^\infty [\Upsilon(\mu_k^{(p)})]^n \, c_k^{(p)}(\x_0).
\end{equation}
Even though this expression remains rather formal, it constitutes the
main result of the paper that lays the theoretical ground for further
analysis of the statistics of adsorptions.  In fact, it incorporates
the effect of the geometrical structure of the confining domain
$\Omega$ through the Steklov eigenvalues and eigenfunctions, and the
adsorption mechanism through the moment-generating function
$\Upsilon(\mu)$ of the threshold $\hat{\ell}$.  When the molecule
has a random lifetime $\tau$ obeying the exponential distribution with
the rate $p$, Eq. (\ref{eq:Qn_tilde}) determines the statistics of
adsorption events realized during the lifetime, $\N_\tau$, in a rather
explicit form.  In turn, the inverse Laplace transform of
Eq. (\ref{eq:Qn_tilde}) is needed to access the statistics
$Q_n(t|\x_0)$ of $\N_t$.  Its analytical inversion, which may involve
the residue theorem, is in general a difficult task.

Moreover, one can deduce the integer-order moments of the number of
adsorptions.  For instance, the mean number of adsorption events,
\begin{equation}
N_1(t|\x_0) = \E_{\x_0}\{ \N_t\} = \sum\limits_{n=1}^\infty n \, Q_n(t|\x_0), 
\end{equation}
reads in the Laplace domain as
\begin{equation}   \label{eq:Np}
\tilde{N}_1(p|\x_0) = \frac{1}{p} \sum\limits_{k=0}^\infty c_k^{(p)}(\x_0) \biggl[\frac{1}{\Upsilon(\mu_k^{(p)})} - 1\biggr]^{-1}  .
\end{equation}
Similarly, one gets the second moment
\begin{align}  \label{eq:Np2}
\tilde{N}_2(p|\x_0) 
& = \tilde{N}_1(p|\x_0) + \frac{2}{p} \sum\limits_{k=0}^\infty c_k^{(p)}(\x_0)
\biggl[\frac{1}{\Upsilon(\mu_k^{(p)})} - 1\biggr]^{-2}  .
\end{align}

As stated earlier, the above analysis assumes that the adsorbed
molecule is immediately released.  It is easy to incorporate a random
duration $\delta_k$ in the adsorbed state.  Let $\phi(t)$ denote the
probability density function of random durations $\delta_1,\ldots,
\delta_n$, which are assumed to be independent from each other and
from the diffusive dynamics.  In this case, $n$ additional
convolutions with $\phi(t)$ should be included into Eq. (\ref{eq:Pnt})
that results in the additional factor $[\tilde{\phi}(p)]^n$ in
Eq. (\ref{eq:Qn_tilde}).  The presence of additional waiting times in
the adsorbed states delays the growth of the number of absorption
events with $t$.  While this effect can be important in practice, its
implementation is straightforward and thus omitted here; in other
words, we restrict our analysis to infinitely short absorption events
and thus ignore $\tilde{\phi}(p)$ in the following analysis.

It is also instructive to recall that the Laplace transform of the
mean boundary local time, $L(t|\x_0) = \E_{\x_0}\{\ell_t\}$, admits a
similar but much simpler form%
\footnote{
Note that $L(t|\x_0)$ can be written explicitly as
\cite{Grebenkov19b}
\begin{equation*}
L(t|\x_0) = D \int\limits_0^t dt' \, \int\limits_{\Gamma} d\x \, G_0(\x,t'|\x_0) ,
\end{equation*}
where $G_0(\x,t'|\x_0)$ is the heat kernel (or the propagator) inside
$\Omega$ with a reflecting boundary $\pa$, satisfying
\begin{equation*}
\partial_t G_0(\x,t|\x_0) = D \Delta G(\x,t|\x_0) ~~ \textrm{in}~\Omega, \qquad
\partial_n G_0(\x,t|\x_0) = 0 ~~ \textrm{on}~ \pa, \qquad
G_0(\x,0|\x_0) = \delta(\x-\x_0).
\end{equation*}
}
\cite{Grebenkov19b}
\begin{equation}   \label{eq:Lp}
\tilde{L}(p|\x_0)  = \frac{1}{p} \sum\limits_{k=0}^\infty \frac{c_k^{(p)}(\x_0)}{\mu_k^{(p)}} \, . 
\end{equation}
Comparing Eqs. (\ref{eq:Np}, \ref{eq:Lp}), one sees how the
moment-generating function $\Upsilon(\mu)$ of the threshold
$\hat{\ell}$ affects each term of the spectral expansion in
Eq. (\ref{eq:Lp}) to transform the mean boundary local time (i.e., the
mean number of adsorption attempts) into the mean number of
adsorptions (i.e., successful outcomes of these attempts).  One may
naturally wonder under which condition these two quantities are
proportional to each other, i.e., a prescribed fraction of adsorption
attempts is successful?  Setting the proportionality condition
\begin{equation}  \label{eq:Nt_Lt}
N_1(t|\x_0) = q L(t|\x_0) 
\end{equation}
with a parameter $q > 0$, one gets immediately
\begin{equation}  \label{eq:Upsilon_exp}
\Upsilon(\mu_k^{(p)}) = \frac{1}{1 + \mu_k^{(p)}/q} 
\end{equation}
for all $\mu_k^{(p)}$, i.e., one gets the exponential law $\Psi(\ell)
= e^{-q\ell}$ for the random threshold $\hat{\ell}$.  In other words,
the exponential law of $\hat{\ell}$ is the necessary and sufficient
condition for getting a constant fraction of successful adsorption
events among all attempts.  As discussed in \cite{Grebenkov20}, this
setting of a constant reactivity $\kappa = qD$ was usually
characterized by the Robin boundary condition on the target.  

The proportionality relation (\ref{eq:Nt_Lt}) can also be seen as a
macroscopic definition of the constant reactivity $\kappa$ via the
ratio between the mean number of adsorptions and the mean boundary
local time time (multiplied by $D$).  This relation generalizes a
recent result by Kay and Giuggioli, which was derived for permeation
dynamics of one-dimensional diffusion on the real line \cite{Kay24}.
They suggested to consider the ratio between the mean number of
permeation events and the mean boundary local time as a macroscopic
definition of permeability.
%
We make a step further and introduce the effective macroscopic
reactivity of the target $\Gamma$ for an arbitrary adsorption kinetics
as
\begin{equation}  \label{eq:kappa_def}
\kappa(t|\x_0) = D \frac{N_1(t|\x_0)}{L(t|\x_0)} \,.
\end{equation}
When the threshold $\hat{\ell}$ exhibits an exponential distribution,
this definition yields a constant reactivity $\kappa$ according to
Eq. (\ref{eq:Nt_Lt}).  However, when the probability law $\Psi(\ell)$
of the threshold $\hat{\ell}$ is not exponential, the ratio in
Eq. (\ref{eq:kappa_def}) may exhibit a nontrivial dependence on both
$t$ and $\x_0$, as illustrated in Sec. \ref{sec:examples}.

\subsection{Long-time behavior for bounded domains}
\label{sec:long}

The inversion of the Laplace transform can be performed via the
residue theorem by identifying the poles of the function
$\tilde{Q}_n(p|\x_0)$ in the complex plane $p\in \C$.  However, this
analysis is rather subtle and requires the knowledge of the Steklov
eigenvalues and eigenfunctions for a given confining domain $\Omega$.
For the sake of clarify, we skip this analysis and focus on the
long-time behavior of the statistics of adsorption events, which
corresponds to the limit $p\to 0$ in the Laplace domain.  This
behavior strongly depends on whether the confining domain $\Omega$ is
bounded or not.  In this section, we consider bounded domains, while
the case of unbounded domains is briefly discussed in
Sec. \ref{sec:unbounded}.

When $\Omega$ is bounded, the smallest eigenvalue $\mu_0^{(p)}$
approaches $\mu_0^{(0)} = 0$, while the corresponding eigenfunction
$V_0^{(p)}$ approaches $V_0^{(0)} = 1/\sqrt{|\Gamma|}$, where
$|\Gamma|$ is the surface area of the adsorbing boundary $\Gamma$.  In
fact, one has \cite{Friedlander91,Grebenkov19b}
\begin{equation}  \label{eq:mu0p}
\mu_0^{(p)} \approx \frac{p|\Omega|}{D|\Gamma|} + O(p^2)   \qquad (p\to 0),
\end{equation} 
where $|\Omega|$ is the volume of the confining domain.  In turn, all
other eigenvalues $\mu_k^{(p)}$ approach their strictly positive
limits $\mu_k^{(0)} > 0$ for $k = 1,2,\ldots$.  Moreover, since the
eigenfunctions $v_k^{(0)}$ are orthogonal to $v_0^{(0)} =
1/\sqrt{|\Gamma|}$, one has
\begin{equation}  \label{eq:ck_asympt_p0}
c_k^{(p)}(\x_0) = \delta_{k,0} + O(p)   \quad (p\to 0),
\end{equation}
where $\delta_{j,k}$ is the Kronecker symbol.

Next, we consider a general asymptotic behavior of the
moment-generating function,
\begin{equation}  \label{eq:Upsilon_p0}
\Upsilon(\mu) \approx 1 - (\mu \ell_0)^\alpha + o(\mu^\alpha)  \qquad (\mu\to 0),
\end{equation}
with a length scale $\ell_0$ and an exponent $0 < \alpha \leq 1$.
Since the mean threshold can be obtained from the moment-generating
function as $\E\{\hat{\ell}\} = - \lim\nolimits_{\mu\to 0}
\partial_\mu \Upsilon(\mu)$, the exponent $\alpha$ distinguishes
two cases when the mean threshold is finite ($\alpha = 1$) or infinite
($\alpha < 1$).

\subsubsection{Mean and standard deviation}
\label{sec:mean_bounded}

Substituting Eqs. (\ref{eq:mu0p}, \ref{eq:Upsilon_p0}) into the first
term in Eqs. (\ref{eq:Np}, \ref{eq:Np2}) and evaluating the
leading-order terms, we get the long-time behavior for any $0 < \alpha
\leq 1$:
\begin{subequations}
\begin{align}     \label{eq:Nt_kappa_long}
N_1(t|\x_0) & \approx \frac{(t/t_0)^{\alpha}}{\Gamma(1+\alpha)} + O(1) , \\
N_2(t|\x_0) & \approx \frac{2(t/t_0)^{2\alpha}}{\Gamma(1+2\alpha)} + O((t/t_0)^\alpha),
\end{align}
\end{subequations}
where $t_0 = \ell_0 |\Omega|/(D|\Gamma|)$.  As a consequence, the
standard deviation of $\N_t$, defined as $\Delta \N_t =
\sqrt{N_2(t|\x_0) - [N_1(t|\x_0)]^2}$, also exhibits a power-law
growth at long times.  However, there is a significant difference
between cases $\alpha < 1$ and $\alpha = 1$.  In the former case, one
gets 
\begin{equation}
\Delta \N_t \approx \sqrt{\frac{2}{\Gamma(1+2\alpha)} - \frac{1}{\Gamma^2(1+\alpha)}}\, (t/t_0)^\alpha \qquad (t\to \infty), 
\end{equation}
i.e., the standard deviation of $\N_t$ grows in the same way as its
mean value.  In other words, the coefficient of variation, $\Delta
N_t/N_1(t|\x_0)$, approaches a constant at long times.  In contrast,
if $\alpha = 1$, the coefficient in front of $(t/t_0)^\alpha$
vanishes, and the standard deviation is determined by the next-order
term, yielding $\Delta \N_t \propto (t/t_0)^{1/2}$.  In this case, the
coefficient of variation vanishes, i.e., the statistics of adsorption
events becomes more concentrated around the mean value.

On the other hand, the mean boundary local time exhibits the long-time
behavior \cite{Grebenkov19b}:
\begin{equation}
L(t|\x_0) \approx \frac{Dt |\Gamma|}{|\Omega|}  \quad (t\to \infty).
\end{equation}
Substituting these asymptotic relations into Eq. (\ref{eq:kappa_def}),
we get the following long-time behavior of the effective reactivity:
\begin{equation}  \label{eq:kappa_long}
\kappa(t|\x_0) \approx \frac{D}{\ell_0} \, \frac{(t/t_0)^{\alpha-1}}{\Gamma(1+\alpha)}  \quad (t\to\infty).
\end{equation}
This relation further clarifies the effect of the adsorption mechanism
through the probability law of the random threshold $\hat{\ell}$:

(i) If the mean threshold $\E\{\hat{\ell}\}$ is finite (and thus equal
to $\ell_0$), one has $\alpha = 1$ and thus the effective reactivity
reaches a constant value $D/\ell_0$.  As a consequence, the effects of
a non-exponential threshold can only emerge at short and intermediate
times, at which $\kappa(t|\x_0)$ may exhibit nontrivial dependence on
$t$ and $\x_0$, but the long-time asymptotic behavior remains similar
to the conventional setting with the exponential law (with the only
difference that the reactivity parameter $q$ is replaced by
$1/\ell_0$).

(ii) In contrast, if the mean threshold $\E\{\hat{\ell}\}$ is infinite
(and thus $\alpha < 1$), the effective reactivity asymptotically
vanishes with time.  In fact, a heavy-tailed distribution of the
threshold allows to produce anomalously large values of $\hat{\ell}$
that require long times to achieve the adsorption condition $\ell_t >
\hat{\ell}$.  Moreover, as the adsorption events are repeated many
times in the limit $t\to\infty$, larger and larger thresholds
$\hat{\ell}_n$ can be generated and yield longer and longer
first-crossing times $\T_n$ so that the fraction of successful
adsorptions decreases with time.  The effect of such non-Markovian
binding mechanism onto diffusion-controlled reactions was discussed in
\cite{Grebenkov23a}.

\subsubsection{Distribution for the case $\alpha < 1$}

Separating the first term with $k = 0$ in Eq. (\ref{eq:Rtilde_def})
from the other terms and substituting Eqs. (\ref{eq:mu0p},
\ref{eq:Upsilon_p0}) into this first term, we get in the limit $p\to
0$
\begin{equation}
\tilde{R}_n(p|\x_0) \approx \frac{1}{p}  \bigl(1 - (pt_0)^\alpha \bigr)^n   ,
\end{equation}
where we used Eq. (\ref{eq:ck_asympt_p0}) to cancel all the terms with
$k > 0$.  Substituting this expression into Eq. (\ref{eq:Qn_tilde})
and applying the inverse Laplace transform, we obtain to the leading
order for any $0 < \alpha < 1$:
\begin{equation}  \label{eq:Qn_asympt}
Q_n(t|\x_0) \approx \frac{(t/t_0)^{-\alpha}}{\Gamma(1-\alpha)}  \qquad (t\to \infty)
\end{equation}
for any fixed $n = 0,1,2,\ldots$.  One sees that the probability of
having exactly $n$ adsorption events slowly vanishes with time.  Note
that getting the opposite limit $n\to \infty$ for a fixed $t$ is a
more difficult task (see an example in Sec. \ref{sec:line}), and these
two limits are not interchangeable.

\subsubsection{Distribution for the case $\alpha = 1$}

A naive extension of the above analysis to the case $\alpha = 1$ is
not valid.  We therefore sketch a different approach based on the
Laplace transform inversion via the Bromwich integral.  Indeed, the
long-time asymptotic behavior of $Q_n(t|\x_0)$ is determined by the
pole $p_0 < 0$ of $\tilde{Q}_n(p|\x_0)$ with the smallest absolute
value $|p_0|$.  The asymptotic behavior would thus be exponential in
the leading order, i.e., $Q_n(t|\x_0) \propto e^{p_0 t}$, with the
prefactor being a polynomial of $t$, which depends on $n$.  The value
of $p_0$ and the form of the prefactor strongly depend on the
moment-generating function $\Upsilon(\mu)$ and the Steklov eigenvalues
$\mu_k^{(p)}$.  Its detailed analysis is beyond the scope of the
paper.

\subsection{Long-time behavior for unbounded domains}
\label{sec:unbounded}

In many situations, one deals with a compact target in the space
$\R^d$, i.e., the confining domain $\Omega = \R^d \backslash
\Omega_0$ is the exterior of a compact set $\Omega_0$.  Since the
confining domain is now unbounded, conventional spectral expansions
over the eigenmodes of the Laplace operator are not suitable.  In
contrast, one can still employ the Steklov eigenmodes since the
underlying eigenvalue problem (\ref{eq:Steklov}) has a discrete
spectrum.  The asymptotic analysis is more subtle and strongly depends
on the space dimension $d$.  Indeed, the one-dimensional case (a
half-line) is elementary (see Sec. \ref{sec:line}) and presents a rare
example when most results can be derived in an explicit form.  The
distribution of the boundary local time for planar domains was studied
in \cite{Grebenkov21a}.  While the developed tools can potentially be
applied to analyze the statistics of adsorption events, its thorough
discussion is beyond the scope of this paper.

In turn, the long-time asymptotic analysis of the three-dimensional
setting, which is the most common for applications, is relatively
simple.  Indeed, as the principal eigenvalue $\mu_0^{(p)}$ approaches
a strictly positive constant $\mu_0^{(0)} > 0$ \cite{Grebenkov24c},
the spectral expansions (\ref{eq:Qn_tilde}) approach constant limits:
\begin{subequations}
\begin{align}
Q_0(\infty|\x_0) & = 1 - \sum\limits_{k=0}^\infty \Upsilon(\mu_k^{(0)}) \, c_k^{(0)}(\x_0) ,\\
Q_n(\infty|\x_0) & = \sum\limits_{k=0}^\infty (1 - \Upsilon(\mu_k^{(0)})) [\Upsilon(\mu_k^{(0)})]^n \, c_k^{(0)}(\x_0) .
\end{align}
\end{subequations}
This asymptotic behavior is the consequence of the transient character
of diffusion in three dimensions.  Indeed, there is a strictly
positive probability for the molecule to escape to infinity, in which
case the statistics of adsorption events remains unchanged.  We also
stress that Eq. (\ref{eq:ck_asympt_p0}), which we used for bounded
domains, does not hold so that (infinitely) many terms can in general
contribute to this statistics.  In this light, the usual case of a
spherical target, for which only the principal eigenmode contributes
(see Sec. \ref{sec:sphere}), is an exception, not a rule.  For
instance, when the target has a spheroidal anisotropic shape,
infinitely many terms do contribute, even though their relative
contributions are progressive attenuated due to increasing eigenvalues
$\mu_k^{(0)}$ \cite{Grebenkov24a}.  Further analysis of this setting
presents an interesting perspective of this work.

\section{Statistics of permeation events}
\label{sec:permeation}

We now briefly discuss a related problem of finding the statistics of
boundary crossings in the more challenging problem of a permeable
interface between two confining domains.  Bressloff extensively
applied the encounter-based approach to describe this phenomenon
\cite{Bressloff22a,Bressloff22c,Bressloff23a,Bressloff23b}.
Let us consider two domains $\Omega_1$ and $\Omega_2$ with a common
interface $\Gamma$.  The molecule starts from a point
$\x_0\in\Omega_1$ and diffuses in $\Omega_1$ until an adsorption on
$\Gamma$, as in the earlier considered case.  However, the adsorbed
molecule is now released on the other side of the interface $\Gamma$,
namely, in $\Omega_2$.  The diffusion continues until an adsorption on
$\Gamma$ and the consequent immediate release in $\Omega_1$, and so
on.  In this way, one can describe successive passages across the
permeable interface $\Gamma$.  As previously, we characterize the
adsorption mechanisms on both sides via random thresholds which are
now governed by two probability laws $\Psi_1(\ell)$ and
$\Psi_2(\ell)$.  One can therefore proceed in constructing the
probability flux densities $j_1(\x,t|\x_0)$ and $j_2(\x,t|\x_0)$ in
both domains.  The number of permeation events $\N_t$ can be defined
as earlier, while its probability distribution is constructed in a
similar way, i.e.,
\begin{align}
Q_0(t|\x_0) & = S_1(t|\x_0) , \\  
Q_{2n}(t|\x_0) & = \int\limits_{\Gamma} d\x_1 \int\limits_0^t dt_1 \, j_1(\x_1,t_1|\x_0) 
\int\limits_{\Gamma} d\x_2 \int\limits_{t_1}^t dt_2 \, j_2(\x_2,t_2-t_1|\x_1) \ldots \\ \nonumber
& \int\limits_{\Gamma} d\x_{2n}  \int\limits_{t_{2n-1}}^t dt_{2n}  
 j_2(\x_{2n},t_{2n}-t_{2n-1}|\x_{2n-1}) S_1(t-t_{2n}|\x_{2n}), \\
Q_{2n+1}(t|\x_0) & = \int\limits_{\Gamma} d\x_1 \int\limits_0^t dt_1 \, j_1(\x_1,t_1|\x_0) 
\int\limits_{\Gamma} d\x_2 \int\limits_{t_1}^t dt_2 \, j_2(\x_2,t_2-t_1|\x_1) \ldots  \\ \nonumber
&\int\limits_{\Gamma} d\x_{2n+1} \int\limits_{t_{2n}}^t dt_{2n+1}  
j_1(\x_{2n+1},t_{2n+1}-t_{2n}|\x_{2n}) S_2(t-t_{2n+1}|\x_{2n+1}),
\end{align}
where we distinguished even and odd crossing numbers.  As previously,
the Laplace transform reduces convolutions in time, whereas spectral
expansions over Steklov eigenfunctions allow one to deal with
convolutions in space.  However, the main difficulty here is that the
Steklov eigenpairs, denoted as $\{\mu_k^{(p,1)}, V_k^{(p,1)}\}$ and
$\{\mu_k^{(p,2)}, V_k^{(p,2)}\}$ in two domains, are different.  As a
consequence, the expressions for $\tilde{Q}_n(p|\x_0)$ become more
sophisticated, e.g.,
\begin{align*}
\tilde{Q}_{2n}(p|\x_0) & = \frac{1}{p}\sum\limits_{k_1,k_2,\ldots,k_{2n},k_{2n+1}}  [V_{k_1}^{(p,1)}(\x_0)]^* \Upsilon_1(\mu_{k_1}^{(p,1)}) 
W_{k_1,k_2} \Upsilon_2(\mu_{k_2}^{(p,2)}) W_{k_2,k_3}^\dagger \ldots \\ \nonumber
& \times W_{k_{2n-1},k_{2n}}^\dagger \Upsilon_2(\mu_{k_{2n}}^{(p,2)}) \biggl(\delta_{k_{2n},k_{2n+1}} 
- W_{k_{2n},k_{2n+1}} \Upsilon_1(\mu_{k_{2n+1}}^{(p,1)})\biggr) \int\limits_{\Gamma} v_{k_{2n+1}}^{(p,1)} ,
\end{align*}
where
\begin{equation}
W_{k,k'} = \int\limits_{\Gamma} v_k^{(p,1)} \, [v_{k'}^{(p,2)}]^* 
\end{equation}
(a similar expression holds for $\tilde{Q}_{2n+1}(p|\x_0)$, while
$\tilde{Q}_0(p|\x_0)$ is equal to $\tilde{S}_1(p|\x_0)$).  The
matrix structure of the above expression allows for a rapid numerical
evaluation of $\tilde{Q}_{2n}(p|\x_0)$ but its theoretical analysis
becomes challenging.

\subsection{Spherical domains}

A considerable simplification appears in the special case when the two
sets of eigenfunctions $\{v_k^{(p,1)}\}$ and $\{v_k^{(p,2)}\}$ on the
target $\Gamma$ are identical.  For instance, if $\Omega_1$ is a ball
of radius $R$ and $\Omega_2$ is a shell between two concentric spheres
of radii $R$ and $L$, the eigenfunctions $v_k^{(p,1)}$ and
$v_k^{(p,2)}$ are equal to spherical harmonics on the spherical
interface $\Gamma$ between two domains.  In this case, $W$ is the
identity matrix, and the above expressions greatly simplify as
\begin{subequations}
\begin{align}
\tilde{Q}_0(p|\x_0) & = \frac{1}{p} - \frac{1}{p} \sum\limits_{k=0}^\infty \Upsilon_1(\mu_k^{(p,1)}) c_k^{(p)}(\x_0)  , \\  
\tilde{Q}_{2n}(p|\x_0) & = \frac{1}{p}\sum\limits_{k=0}^\infty  
\bigl[\Upsilon_1(\mu_k^{(p,1)}) \Upsilon_2(\mu_k^{(p,2)})\bigr]^n 
\biggl(1 - \Upsilon_1(\mu_k^{(p,1)})\biggr) c_{k}^{(p)}(\x_0)  , \\ 
\tilde{Q}_{2n+1}(p|\x_0) & = \frac{1}{p}\sum\limits_{k=0}^\infty  
\bigl[\Upsilon_1(\mu_k^{(p,1)}) \Upsilon_2(\mu_k^{(p,2)})\bigr]^n \Upsilon_1(\mu_k^{(p,1)}) 
\biggl(1 - \Upsilon_2(\mu_k^{(p,2)})\biggr) c_{k}^{(p)}(\x_0)   ,
\end{align}
\end{subequations}
where
\begin{align}
c_k^{(p)}(\x_0) & = [V_k^{(p,1)}(\x_0)]^* \int\limits_{\Gamma} v_k^{(p,1)} .
\end{align}
A lengthy but elementary calculation yields the mean number of
permeation events in the Laplace domain
\begin{equation}
\tilde{N}_1(p|\x_0) = \frac{1}{p} \sum\limits_{k=0}^\infty \frac{\Upsilon_1(\mu_k^{(p,1)}) [1 +\Upsilon_2(\mu_k^{(p,2)})]}
{1 - \Upsilon_1(\mu_k^{(p,1)}) \Upsilon_2(\mu_k^{(p,2)})}  c_k^{(p)}(\x_0).
\end{equation}
If the domains and adsorptions mechanisms were identical, i.e., if
$\mu_k^{(p,1)} = \mu_k^{(p,2)} = \mu_k^{(p)}$ and $\Upsilon_1(\mu) =
\Upsilon_2(\mu) = \Upsilon(\mu)$, the above expression would reduce to
Eq. (\ref{eq:Np}), as it should.

As the long-time behavior corresponds to the limit $p\to 0$, one needs
to distinguish bounded and unbounded domains.  In the former case, one
has $\mu_0^{(p,i)} \approx p |\Omega_i|/(D|\Gamma|)$.  Substituting
this expression into 
\begin{equation}
\Upsilon_i(\mu) = 1 - (\mu \ell_i)^{\alpha_i} + o(\mu^\alpha_i) \qquad (\mu\to 0),
\end{equation}
we get in the leading order:
\begin{equation}
\tilde{N}_1(p|\x_0) \approx \frac{2}{p[(p t_1)^{\alpha_1} + (p t_2)^{\alpha_2}]}  \qquad (p\to 0) \,,
\end{equation}
where $t_i = \ell_i |\Omega_i|/(D|\Gamma|)$.  If $\alpha_1 = \alpha_2
= \alpha$, one finds
\begin{equation}
N_1(t|\x_0) \approx \frac{2 t^{\alpha}}{\Gamma(1+\alpha)[t_1^{\alpha} +t_2^{\alpha}]} \qquad (t\to \infty).
\end{equation}
If $\alpha_1 < \alpha_2$, the slower adsorption mechanism from
$\Omega_1$ on $\Gamma$ determines the long-time asymptotic behavior:
\begin{equation}
N_1(t|\x_0) \approx \frac{2 t^{\alpha_1}}{\Gamma(1+\alpha_1) t_1^{\alpha_1}} \qquad (t\to \infty).
\end{equation}

In turn, if one of two domains is unbounded (say, $\Omega_2$), all
eigenvalues $\mu_k^{(p,2)}$ approach to strictly positive limits
$\mu_k^{(0,2)}$ in three dimensions so that the leading term is
different, $\tilde{N}_1(p|\x_0) = O(1/p)$, yielding another long-time
asymptotic behavior
\begin{equation}
N_1(\infty|\x_0) =  \sum\limits_{k=0}^\infty \frac{\Upsilon_1(\mu_k^{(0,1)}) [1 +\Upsilon_2(\mu_k^{(0,2)})]}
{1 - \Upsilon_1(\mu_k^{(0,1)}) \Upsilon_2(\mu_k^{(0,2)})}  c_k^{(0)}(\x_0) .
\end{equation}
The constant limit of $N_1(t|\x_0)$ reflects the transient diffusion
in the unbounded domain $\Omega_2$, i.e., an eventual escape of the
molecule to infinity.

\section{Two relevant examples}
\label{sec:examples}

While our formalism allows one to investigate the statistics of
adsorption and permeation events in arbitrary environments, its
application requires a numerical computation of the Steklov eigenmodes
and inverse Laplace transform.  Even though various numerical methods
are available (see \cite{Chaigneau24} and references therein), we
focus here on two practically relevant examples: a flat surface
(Sec. \ref{sec:line}) and a spherical surface (Sec. \ref{sec:sphere}).
In both cases, the Steklov eigenmodes are known explicitly that yields
closed expressions for the statistics.  It is worth noting that the
overwhelming majority of earlier studies of adsorption phenomena and
reversible binding reactions were also limited to these two settings
(see \cite{Scher23,Grebenkov23a} and references therein).

\subsection{Flat surface}
\label{sec:line}

We first consider molecular diffusion in the upper half-space bounded
by an adsorbing horizontal plane.  As lateral displacements of the
molecule do not affect adsorption or permeation events, this problem
is equivalent to diffusion on the positive half-line, $\Omega =
(0,+\infty)$, for which the Steklov problem has a single eigenpair,
\begin{equation}
\mu_0^{(p)} = \sqrt{p/D},  \qquad V_0^{(p)}(x) = e^{-x \sqrt{p/D}} ,
\end{equation}
so that $c_0^{(p)}(x_0) = e^{-x_0\sqrt{p/D}}$ from
Eq. (\ref{eq:ck_def}) and thus
\begin{subequations}  \label{eq:Qn_tilde_halfline}
\begin{align}
\tilde{Q}_0(p|x_0) & = \frac{1}{p} - \frac{e^{-x_0\sqrt{p/D}}}{p}  \Upsilon(\sqrt{p/D})   , \\
\tilde{Q}_n(p|x_0) & = \frac{e^{-x_0\sqrt{p/D}}}{p}  [\Upsilon(\sqrt{p/D})]^n [1-\Upsilon(\sqrt{p/D})].
\end{align}
\end{subequations}

\begin{figure}
\begin{center}
\includegraphics[width=0.49\textwidth]{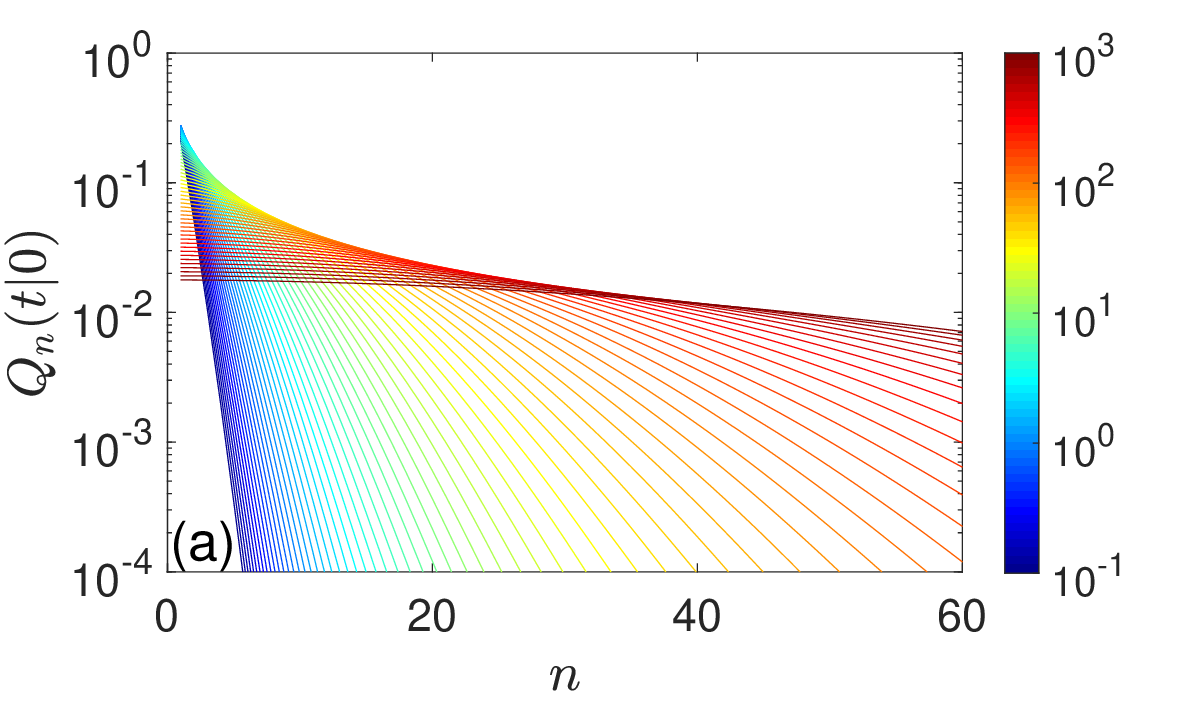} 
\includegraphics[width=0.49\textwidth]{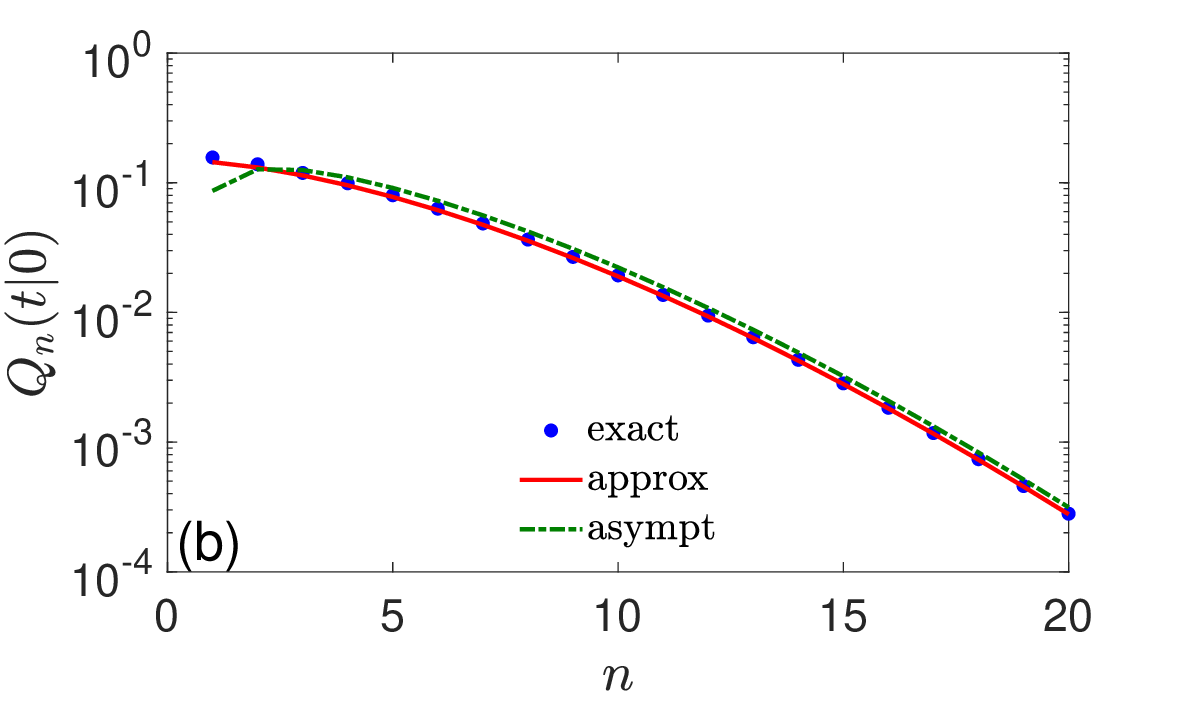} 
\end{center}
\caption{
Statistics of adsorption events at the origin for diffusion on a
half-line with a constant reactivity $q = 1$.  {\bf (a)} $Q_n(t|0)$ as
a function of $n$ for 64 values of $t$, chosen equidistantly on
logarithmic scale, from $t = 10^{-1}$ (dark blue) to $t = 10^3$ (dark
red), with $D = 1$.
{\bf (b)} $Q_n(t|0)$ as a function of $n$ for $t = 10$.  Filled
circles present the exact solution (\ref{eq:Qn0_line_exact}), solid
line shows the approximation (\ref{eq:Qn0_line_approx}), whereas
dash-dotted line indicates the asymptotic relation
(\ref{eq:Qn0_line_asympt}).}
\label{fig:Qn_line}
\end{figure}

For the case of a constant reactivity and $x_0 = 0$, one can
substitute $\Upsilon(\sqrt{p/D}) = 1/(1 + \sqrt{p/D}/q)$ into this
expression and then invert the Laplace transform to represent
$Q_n(t|0)$ in terms of the parabolic cylinder function (see
\cite{Prudnikov5}, p. 43, eq. 21).  After simplifications, we get
\begin{equation}  \label{eq:Qn0_line_exact}
Q_n(t|0) = \frac{1}{\sqrt{\pi}} (q^2 Dt)^{n/2} \, U\bigl((n+1)/2,1/2, q^2 Dt\bigr),
\end{equation}
where $U(a,b,z)$ is the Tricomi confluent hypergeometric function.  We
therefore retrieved the result by Kay and Giuggioli \cite{Kay24}.
Note that $Q_0(t|0) = \erfcx\bigl(q\sqrt{Dt}\bigr)$, where $\erfcx(z)
= e^{z^2} \erfc(z)$ is the scaled complementary error function (note
that $Q_0(t|x_0)$ can also be written explicitly for any $x_0 > 0$).
Figure \ref{fig:Qn_line}(a) illustrates the dependence of $Q_n(t|0)$
of $n$ for a broad range of times.

Since $U(a,b,z) \approx z^{-a}$ as $z\to\infty$, we get
\begin{equation}
Q_n(t|0) \approx \frac{1}{\sqrt{\pi}} (q^2 Dt)^{-1/2}    \qquad (t\to \infty).
\end{equation}  
Curiously, the long-time decay of these probabilities is very slow and
independent of $n$.  

In turn, the opposite limit of large $n$ can be obtained from the
asymptotic behavior of $U(a,b,z)$ for large $a$ (see \cite{formula}) 
that yields
\begin{equation}  \label{eq:Qn0_line_approx}
Q_n(t|0) \approx \frac{2 z^{n/2} e^{z/2 - \beta(n+1)}}{\sqrt{n+1} \, \Gamma((n+1)/2) \sqrt{1 + e^{-w}}} \,,
\end{equation}
where $z = q^2 Dt$, $w = \arccosh(1+z/(n+1))$, and $\beta = (w+\sinh
w)/2$.  While the dependence on $n$ is somewhat cumbersome here, this
asymptotic relation turns out to be remarkably accurate for all $t$
(Fig. \ref{fig:Qn_line}b).  When $n \gg z$, one gets $\beta \approx w
\approx \sqrt{2z/n} \ll 1$ so that
\begin{equation}  \label{eq:Qn0_line_asympt}
Q_n(t|0) \approx \frac{z^{n/2} e^{z/2 - \sqrt{2nz} - \sqrt{2z/n}}}{\sqrt{n/2}~ \Gamma(n/2)} \,, 
\end{equation}
and the factor $\Gamma(n/2)$ provides faster-than-exponential decay of
$Q_n(t|0)$ as $n\to \infty$.

One can also compute the moments of the number of adsorption events;
for instance, the spectral expansion in Eq. (\ref{eq:Np}) contains a
single term, and its inverse Laplace transform yields $N_1(t|x_0) = q
L(t|x_0)$, with
\begin{equation}  \label{eq:Lt_line}
L(t|x_0) = \frac{\sqrt{4Dt}}{\sqrt{\pi}} e^{-x_0^2/(4Dt)} - x_0 \, \erfc\bigl(x_0/\sqrt{4Dt}\bigr).
\end{equation}
At $x_0 = 0$, one gets $N_1(t|0) = 2q \sqrt{Dt}/\sqrt{\pi}$, which is
also the long-time asymptotic behavior for any $x_0 \geq 0$.  This
sublinear growth with time is a consequence of the unboundedness of
the half-line: the molecule can diffuse far away from the origin to
keep the number of adsorption events constant for long periods.  In
contrast, $N_1(t|\x_0)$ grows linearly with time for any bounded
domain at long times, see Eq. (\ref{eq:Nt_kappa_long}).

Since the spectrum of the Steklov problem is particularly simple for a
flat surface, one can determine the statistics of the adsorption
events for any adsorption mechanism defined via a given
moment-generating function $\Upsilon(\mu)$.  To illustrate its effect,
we choose a simple model when the molecule adsorbs after a fixed
amount of attempts.  This deterministic threshold, $\hat{\ell} =
\ell_0$, with some $\ell_0 > 0$, corresponds to $\Upsilon(\mu) =
e^{-\mu \ell_0}$.  Substituting this expression into
Eqs. (\ref{eq:Qn_tilde_halfline}) and performing the inverse Laplace
transform, we get
\begin{subequations}  \label{eq:Qn_tilde_halfline_fixed}
\begin{align}
Q_0(t|x_0) & = 1 - \erfc\biggl(\frac{x_0 + \ell_0}{\sqrt{4Dt}}\biggr)   , \\
Q_n(t|x_0) & = \erfc\biggl(\frac{x_0 + n\ell_0}{\sqrt{4Dt}}\biggr) - \erfc\biggl(\frac{x_0 + (n+1)\ell_0}{\sqrt{4Dt}}\biggr).
\end{align}
\end{subequations}
In turn, the mean number of adsorptions is 
\begin{equation}  \label{eq:Nt_fixed}
N_1(t|x_0) = \sum\limits_{n=1}^\infty \erfc\biggl(\frac{x_0 + n\ell_0}{\sqrt{4Dt}}\biggr).
\end{equation}

\begin{figure}
\begin{center}
\includegraphics[width=0.49\textwidth]{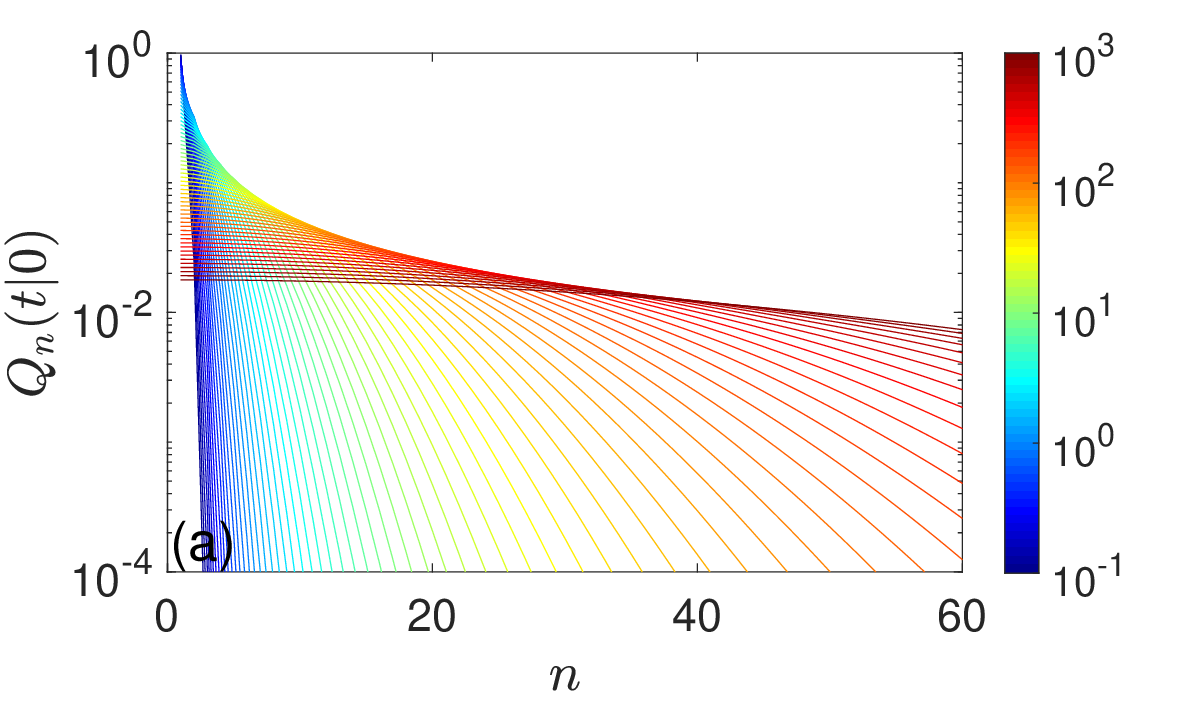} 
\includegraphics[width=0.49\textwidth]{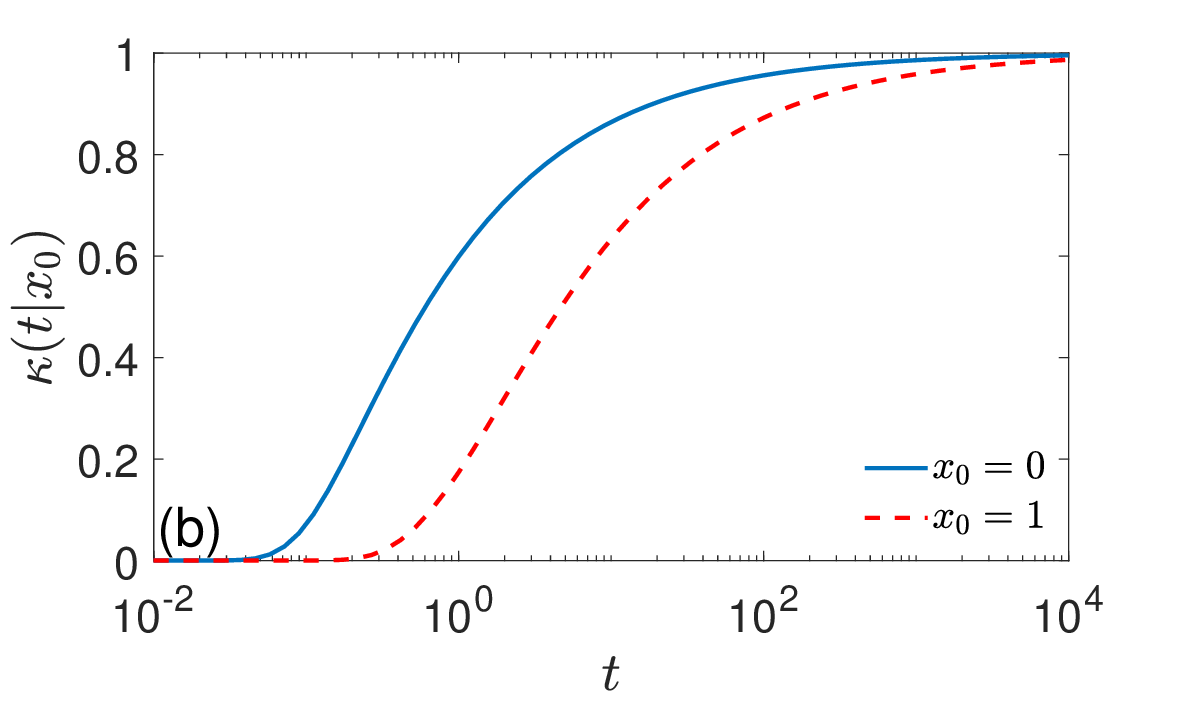} 
\end{center}
\caption{
Statistics of adsorption events at the origin for diffusion on a
half-line for the adsorption mechanism with a fixed threshold $\ell_0
= 1$.  {\bf (a)} $Q_n(t|0)$ as a function of $n$ for 64 values of $t$,
chosen equidistantly on logarithmic scale, from $t = 10^{-1}$ (dark
blue) to $t = 10^3$ (dark red), with $D = 1$.
{\bf (b)} The effective reactivity $\kappa(t|x_0)$ defined in
Eq. (\ref{eq:kappa_def}), with $N_1(t|x_0)$ given by
Eq. (\ref{eq:Nt_fixed}) and $L(t|x_0)$ given by
Eq. (\ref{eq:Lt_line}).  }
\label{fig:Qn_line_fixed}
\end{figure}

Figure \ref{fig:Qn_line_fixed}a illustrates the statistics of
adsorption events for a fixed threshold $\ell_0 = 1$.  It is
instructive to compare this panel with Fig. \ref{fig:Qn_line}a for a
constant reactivity $q = 1$.  Even though the functional forms of
$Q_n(t|0)$ in Eqs. (\ref{eq:Qn_tilde_halfline},
\ref{eq:Qn_tilde_halfline_fixed}) are different, these two
statistics look similar at long times.  This is not surprising as the
exponentially distributed random threshold $\hat{\ell}$ has the mean
value $1/q = 1$, which is identical to the fixed threshold $\ell_0 =
1$.  In turn, some distinctions can be noticed at short times, in
particular, $Q_n(t|0)$ decays faster with $n$ when the threshold is
fixed.  The distinction at short times becomes even more explicit when
plotting the effective reactivity $\kappa(t|x_0)$ defined in
Eq. (\ref{eq:kappa_def}) as the ratio between the mean number of
adsorptions, $N_1(t|x_0)$, and the mean boundary local time
$L(t|x_0)$.  As this ratio is equal to $1$ for a constant reactivity
$\kappa = qD = 1$, its deviations from $1$ witnesses the effect of the
adsorption mechanism.  Since the mean threshold is finite for the
fixed threshold $\hat{\ell} = \ell_0$, the effectively reactivity
$\kappa(t|x_0)$ approaches $D/\ell_0 = 1$ at long times.  In turn,
$\kappa(t|x_0)$ vanishes at short times.  Indeed, as $t\to 0$, the
molecule has almost no chance to hit the boundary enough times to
exceed a fixed threshold $\hat{\ell} = \ell_0 = 1$.  In turn, when the
threshold $\hat{\ell}$ is exponentially distributed, its realized
value can be much smaller, so that the adsorption at short times
becomes more probable.  We note also that if the starting point $x_0$
does not lie on the boundary (i.e., $x_0 > 0$), the molecule needs
first to reach the boundary that shifts the curve $N_1(t|x_0)$ to the
right (to longer times).  However, the starting point $x_0$ has no
effect on the long-time asymptotic behavior, as expected.

Following \cite{Grebenkov23a}, we also consider the Mittag-Leffler
probability law for the threshold $\hat{\ell}$ as a model of anomalous
reactivity:
\begin{equation}  \label{eq:Psi_ML}
\Psi(\ell) = E_{\alpha}(-(\ell/\ell_0)^\alpha)  \quad \Rightarrow \quad \Upsilon(\mu) = \frac{1}{1 + (\mu \ell_0)^\alpha} \,,
\end{equation} 
with a length scale $\ell_0$ and an exponent $0 < \alpha \leq 1$,
where $E_{\alpha}(z) = \sum\nolimits_{n=0}^\infty
z^n/\Gamma(n+\alpha)$ is the Mittag-Leffler function.  At $\alpha =
1$, one retrieves the exponential distribution with the mean threshold
$\ell_0$.  Substitution of Eq. (\ref{eq:Psi_ML}) into Eq. (\ref{eq:Np})
and the inversion of the Laplace transform with $x_0 = 0$ yield the
exact result
\begin{equation}
N_1(t|0) = \frac{(Dt/\ell_0^2)^{\alpha/2}}{\Gamma(1 + \alpha/2)}    \,.
\end{equation}
Comparing this expression with $L(t|0) = 2\sqrt{Dt}/\sqrt{\pi}$, we
get the effective reactivity
\begin{equation}
\kappa(t|0) = D \frac{N_1(t|0)}{L(t|0)} = \frac{(D/\ell_0) \sqrt{\pi}}{2\Gamma(1+\alpha/2)} (Dt/\ell_0^2)^{(\alpha-1)/2} \,.
\end{equation}
We stress that this exact formula is valid for any $t$.  As discussed
earlier in Sec. \ref{sec:long}, the effective reactivity vanishes at
long times.  However, the exponent is twice smaller as compared to the
case of bounded domains, see Eq. (\ref{eq:kappa_long}).  In turn, this
effective reactivity diverges at short times.  This counter-intuitive
behavior is a consequence of the Mittag-Leffler model.  In fact, the
probability density of the threshold $\hat{\ell}$ behaves as
$(-\partial_\ell \Psi(\ell)) \approx (\ell/\ell_0)^{\alpha-1}/(\ell_0
\Gamma(\alpha))$ as $\ell\to 0$.  The divergence of this density
implies that small values of the threshold $\hat{\ell}$ are too
probable so that the adsorption events are very frequent at short
times (see also a related discussion in \cite{Grebenkov23a}).

\subsection{Spherical surface}
\label{sec:sphere}

When $\Omega = \{ \x\in \R^3 ~:~ |\x| < R\}$ is a ball of radius $R$
with the adsorbing boundary $\Gamma = \pa$, the Steklov eigenvalue
problem admits an explicit solution \cite{Levitin,Grebenkov20c}.  In
spherical coordinates $\x = (r,\theta,\phi)$, one has
\begin{equation}
\mu_k^{(p)} = \sqrt{p/D}\, \frac{i'_k(R \sqrt{p/D})}{i_k(R\sqrt{p/D})},  \qquad 
V_k^{(p)} = \frac{i_k(r \sqrt{p/D})}{i_k(R\sqrt{p/D})} Y_{k,m}(\theta,\phi),
\end{equation}
where prime denotes the derivative with respect to the argument,
$i_n(z) = \sqrt{\pi/(2z)} I_{n+1/2}(z)$ are the modified spherical
Bessel functions of the first kind, and $Y_{k,m}(\theta,\phi)$ are the
normalized spherical harmonics.  While the enumeration of spherical
harmonics and thus of Steklov eigenfunctions requires the second index
$m$ (ranging from $-k$ to $k$), it can be omitted in our analysis.
Indeed, since $v_0^{(p)} = Y_{0,0} = 1/\sqrt{4\pi R^2}$, the integrals
of $v_k^{(p)}$ over $\Gamma$ vanish for all $k > 0$ due to
orthogonality of spherical harmonics.  As a consequence, one has
\begin{equation}
c_k^{(p)}(\x_0) = \frac{i_0(r_0 \sqrt{p/D})}{i_0(R\sqrt{p/D})} \delta_{k,0} ,
\end{equation}
where $r_0 = |\x_0|$ is the radial coordinate of the starting point.
We get then an explicit solution in the Laplace domain:
\begin{subequations}
\begin{align}
\tilde{Q}_0(p|r_0) & = \frac{1}{p} - \frac{i_0(r_0 \sqrt{p/D})}{p \,i_0(R\sqrt{p/D})} \Upsilon(\mu_0^{(p)}) ,\\
\tilde{Q}_n(p|r_0) & = \frac{i_0(r_0\sqrt{p/D})}{p \,i_0(R\sqrt{p/D})} [\Upsilon(\mu_0^{(p)})]^n (1 - \Upsilon(\mu_0^{(p)})) ,
\end{align}
\end{subequations}
with $i_0(z) = \sinh(z)/z$.  In turn, the Laplace inversion needs to
be performed numerically for a given moment-generating function
$\Upsilon(\mu)$.

\begin{figure}
\begin{center}
\includegraphics[width=0.49\textwidth]{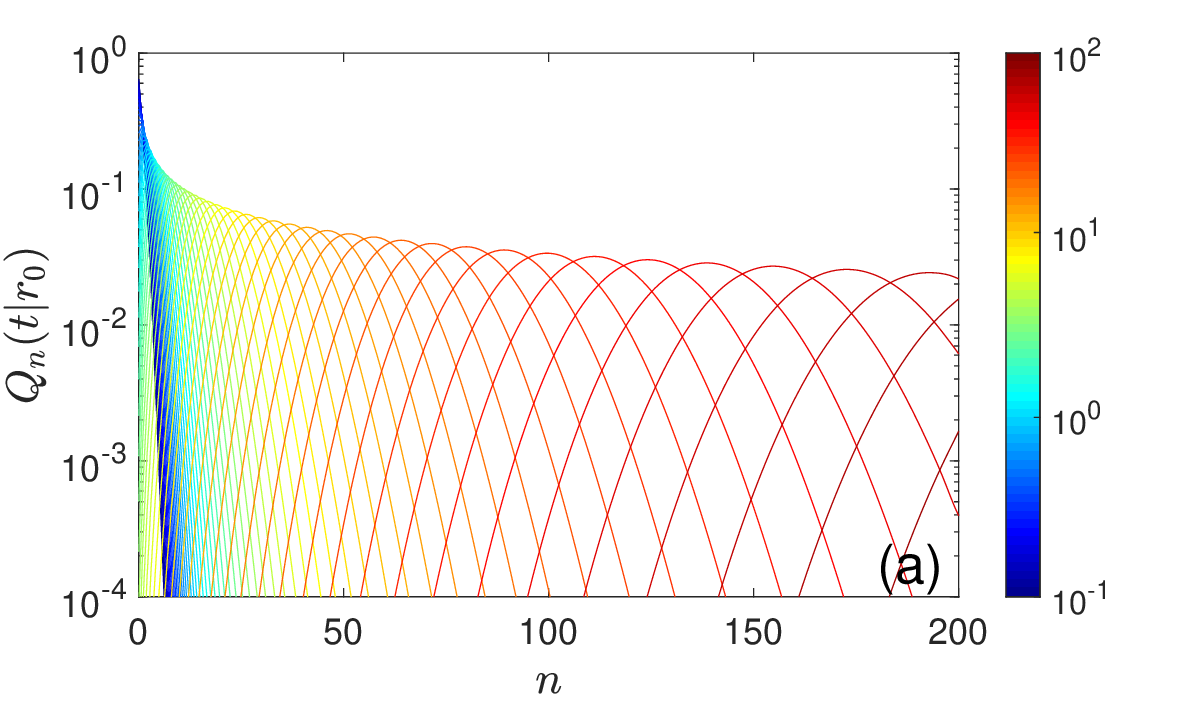} 
\includegraphics[width=0.49\textwidth]{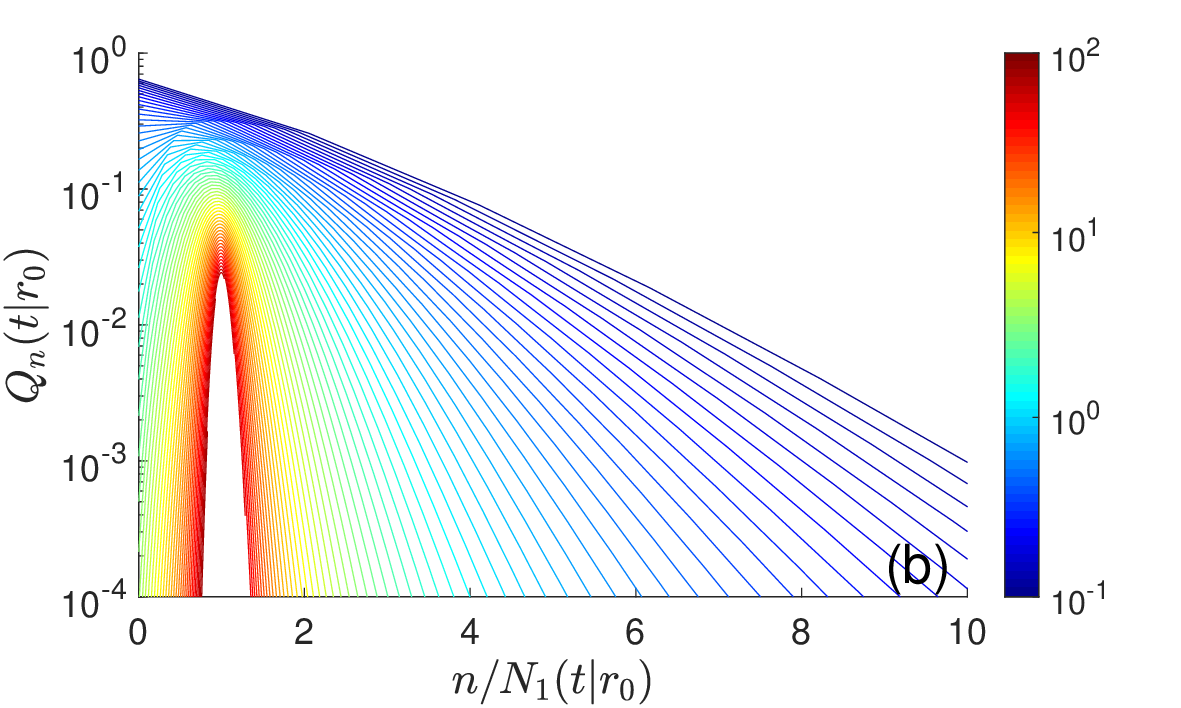} 
\includegraphics[width=0.49\textwidth]{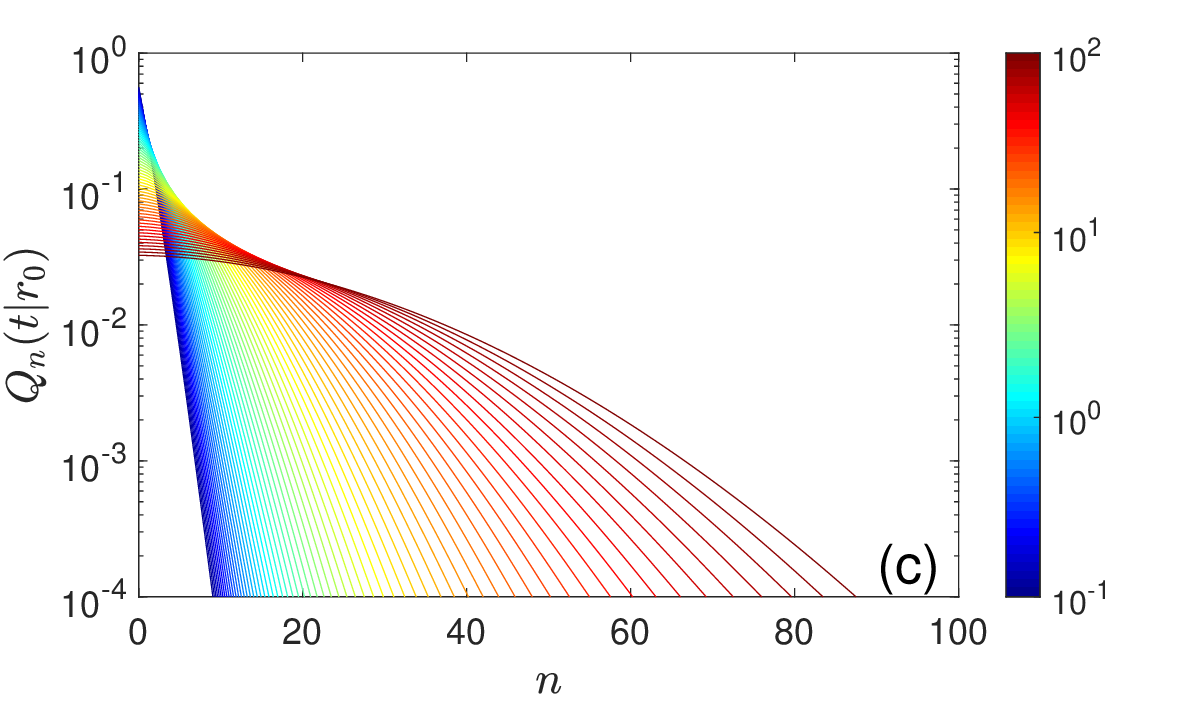} 
\includegraphics[width=0.49\textwidth]{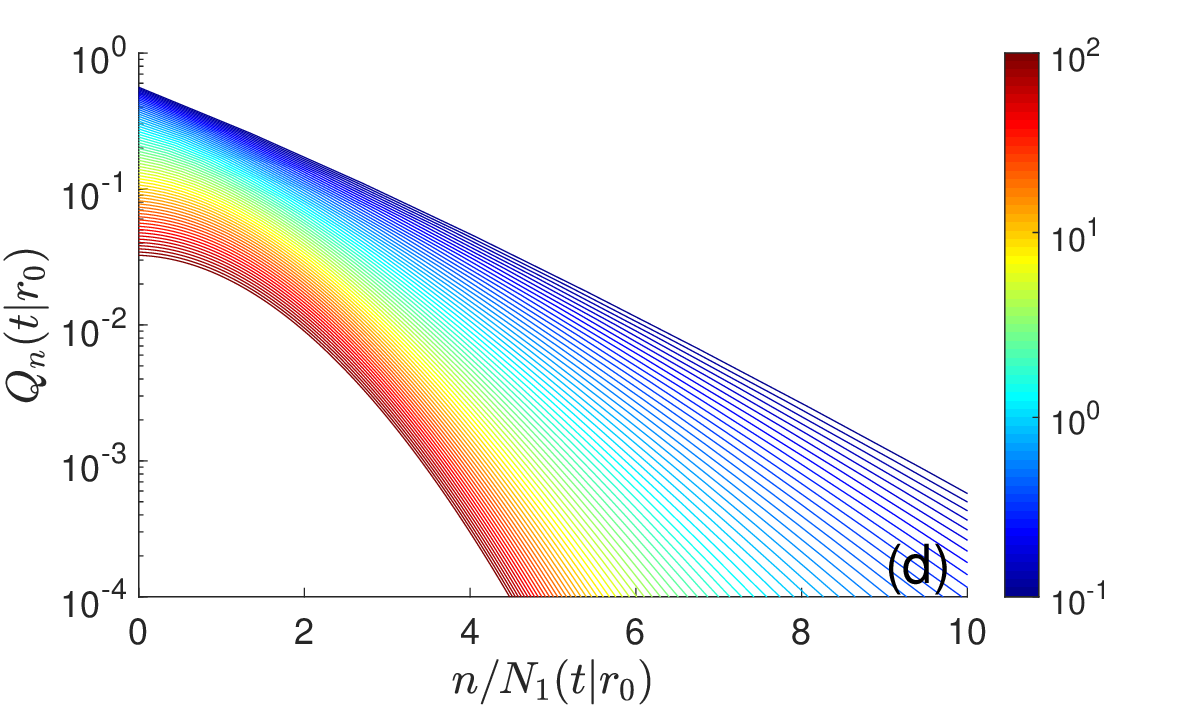} 
\end{center}
\caption{
Statistics of adsorption events on the adsorbing sphere of radius $R =
1$ with a constant reactivity $q = 1$ (top panels) and with the
Mittag-Leffler adsorption model with $\alpha = 0.5$ and $\ell_0 = 1$
(bottom panels). {\bf (a,c)} $Q_n(t|R)$ as a function of $n$ for 64
values of $t$, chosen equidistantly on logarithmic scale, from $t =
10^{-1}$ (dark blue) to $t = 10^2$ (dark red), with $D = 1$.  {\bf
(b,d)} The same probabilities with $n$ rescaled by the mean number of
adsorptions.}
\label{fig:Qn_sphere}
\end{figure}

Figure \ref{fig:Qn_sphere}(a) illustrates the statistics of adsorption
events on the unit sphere with a constant reactivity $q$.  A visual
comparison of this figure with Fig. \ref{fig:Qn_line}(a) highlights
significant differences.  While the dependence of $Q_n(t|x_0)$ on $n$
remained almost flat for diffusion in the half-line, there is a
distinct maximum of $Q_n(t|r_0)$ for each $t$ large enough.  In other
words, at long times, it is highly unlikely to experience a small
number of adsorptions in the case of molecular diffusion inside a
bounded domain.  The maximum is shifted to larger values of $n$ as
time increases.  Moreover, the distribution becomes more and more
concentrated around the mean value, as discussed in
Sec. \ref{sec:mean_bounded}.  To outline this narrowing effect,
Fig. \ref{fig:Qn_sphere}(b) presents the same distribution but plotted
against the rescaled number $n/N_1(t|r_0)$.  As time increases, this
distribution gets narrower, and its width, characterized by the
coefficient of variation, vanishes.

The behavior is different for adsorption mechanisms with infinite mean
threshold $\E\{\hat{\ell}\}$.  Figure \ref{fig:Qn_sphere}(c) shows the
statistics of adsorptions for the Mittag-Leffler model
(\ref{eq:Psi_ML}) with $\alpha = 0.5$.  This panel resembles
Fig. \ref{fig:Qn_line}(a) for diffusion on the half-line.  As shown in
Sec. \ref{sec:mean_bounded}, the coefficient of variation approaches a
constant, i.e., the distribution does not become narrower.

\section{Discussion and conclusion}
\label{sec:conclusion}

In this paper, we developed a general theoretical framework for
accessing the statistics of adsorption and permeation events in
molecular diffusion.  These statistics depend on the geometric shape
of the confinement and the adsorption/permeation mechanism on the
boundary.  The geometry is captured via the Steklov eigenmodes,
whereas the mechanism is introduced via the random threshold
$\hat{\ell}$ and its moment-generating function $\Upsilon(\mu) =
\E\{e^{-\mu \hat{\ell}}\}$.  Using the encounter-based approach, we
derived the spectral expansion (\ref{eq:Qn_tilde}) for the Laplace
transform $\tilde{Q}_n(p|\x_0)$ of the probability $Q_n(t|\x_0)$ of
having $n$ adsorption events up to time $t$ (i.e., of $\N_t = n$).  A
similar, yet much more formal expansion was derived for the statistics
of permeation events.  On one hand, the inverse Laplace transform of
$\tilde{Q}_n(p|\x_0)$ allows one to get $Q_n(t|\x_0)$, at least
numerically.  Moreover, the small-$p$ asymptotic behavior of
$\tilde{Q}_n(p|\x_0)$ determines the long-time asymptotic behavior of
$Q_n(t|\x_0)$, as we illustrated for bounded and unbounded domains.
The long-time statistics strongly depend on the chosen adsorption
mechanism via the asymptotic expansion of the moment-generating
function $\Upsilon(\mu)$, in particular, whether the mean threshold
$\E\{\hat{\ell}\}$ is finite or not.  On the other hand, if the
molecule has an exponentially distributed random lifetime, the Laplace
transform $\tilde{Q}_n(p|\x_0)$ provides directly the statistics of
adsorption events realized prior to degradation, passivation or
irreversible transformation of the molecule.

We also discussed the behavior of the mean number of adsorptions
$N_1(t|\x_0)$, its second moment $N_2(t|\x_0)$, and the resulting
standard deviation.  Moreover, we introduced the effective reactivity
$\kappa(t|\x_0)$ as the ratio between the mean number of adsorptions
and the mean boundary local time $L(t|\x_0)$.  The latter can be
understood as a proxy for the mean number of adsorption attempts,
whereas $N_1(t|\x_0)$ is the mean number of their successful outcomes.
We showed that $\kappa(t|\x_0)$ remains constant in the common setting
of a constant reactivity, which is characterized by the exponential
distribution of the threshold $\hat{\ell}$ or, equivalently, by the
Robin boundary condition.  In general, however, the effective
reactivity is not constant, and its time and space dependence reveals
the impact of the adsorption mechanism.  In particular, at long times,
$\kappa(t|\x_0)$ can either approach a constant or vanish, depending
on the adsorption mechanism.  This behavior is rather universal for
bounded confining domains.  In turn, the short-time behavior is more
sensitive to the particular choice of the moment-generating function
$\Upsilon(\mu)$ and may even depend on the starting point $\x_0$.

The general theoretical framework was illustrated on two practically
relevant examples of a flat surface and a spherical surface.  In the
former case, the Steklov spectrum contains a single, fully explicit
eigenmode that dramatically simplifies the analysis.  In particular,
we managed to obtain $Q_n(t|x_0)$ explicitly for two adsorption
mechanisms: a constant reactivity and a fixed threshold.  For a
spherical boundary, the spectrum of the Steklov problem contains
infinitely many eigenmodes but the rotational symmetry implies that
only the principal eigenmode contributes to the spectral expansion
(\ref{eq:Qn_tilde}).

The progress in optical imaging allows one gather significant amounts
of single-particle trajectories in various biological and soft matter
systems.  The statistics of adsorption and permeation events can thus
potentially be obtained from these data.  One can then formulate a
number of optimization and inverse problems.  For instance, what can
one say about the geometric confinement and/or the
adsorption/permeation mechanism from the observed statistics?  Is it
possible to optimize the shape of the substrate to increase or
decrease the number of adsorption events or to reshape their
distribution?  If such optimal solutions exist, how do they match with
the actual structure of biological and chemical systems?  The
developed formalism provides a mathematical foundation for addressing
these challenging problems in the future.

\begin{acknowledgments}
The author acknowledges the Simons Foundation for supporting his
sabbatical sojourn in 2024 at the CRM (University of Montr\'eal,
Canada), as well as the Alexander von Humboldt Foundation for support
within a Bessel Prize award.
\end{acknowledgments}

\end{document}